\documentclass{article}
\usepackage[T1]{fontenc}
\usepackage[latin9]{inputenc}
\usepackage[letterpaper]{geometry}
\usepackage{textcomp}
\usepackage{amstext}
\usepackage{graphicx}
\usepackage{esint}
\usepackage[authoryear]{natbib}

\makeatletter
\newcommand{\lyxaddress}[1]{
	\par {\raggedright #1
	\vspace{1.4em}
	\noindent\par}
}

\makeatother
\begin{document}

\title{Constraining Jupiter's internal flows using Juno magnetic and gravity
measurements}

\author{Eli Galanti$^{1}$, Hao Cao$^{2}$, and Yohai Kaspi$^{1}$
\\
\\
(Published in GRL: https://doi.org/10.1002/2017GL074903)}
\maketitle

\lyxaddress{\begin{center}
\textit{$^{1}$Department of Earth and Planetary Sciences, Weizmann
Institute of Science, Rehovot, Israel. }\\
\textit{$^{2}$Division of Geological and Planetary Sciences, California Institute of Technology, Pasadena, CA, USA}
\par\end{center}}

\begin{abstract}
Deciphering the flow below the cloud-level of Jupiter remains a critical
milestone in understanding Jupiter's internal structure and dynamics.
The expected high-precision Juno measurements of both the gravity
field and the magnetic field might help to reach this goal. Here we
propose a method that combines both fields to constrain the depth
dependent flow field inside Jupiter. This method is based on a mean-field
electrodynamic balance that relates the flow field to the anomalous
magnetic field, and geostrophic balance that relates the flow field
to the anomalous gravity field. We find that the flow field has two
distinct regions of influence - an upper region in which the flow
affects mostly the gravity field, and a lower region in which the
flow affects mostly the magnetic field. An optimization procedure
allows to reach a unified flow structure that is consistent with both
the gravity and the magnetic fields.
\end{abstract}

\section{Introduction{\normalsize{}\label{sec:Introduction}}}

The winds engulfing Jupiter, manifested in the planet's cloud motion,
have been studied and analyzed extensively, resulting in a robust
spatial picture of their amplitude and direction \citep{Porco2003,Choi2011}.
However, their behavior below the clouds remains largely unknown.
Aside from the single direct measurement of the flow below the clouds
by the Galileo probe at a specific location near $6^{\circ}$N \citep{Atkinson1996},
there is very little knowledge about the nature of the flow underneath
the observable clouds. Starting in August 2016, the Juno spacecraft
began measuring a range of physical parameters including the planetary
magnetic and gravity fields while orbiting Jupiter at unprecedented
proximity \citep{Bolton2017a}. These measurements, once calibrated
and analyzed, will provide unprecedented latitude and longitude dependent
high-precision maps of both fields, which can be used to construct
the depth-dependent flow field of Jupiter. 

Several studies examined the connection between the flow within Jupiter
and the spatial variations in the planetary gravity field. It was
shown that the gravity field measurements could be used to infer the
internal structure of the flow below its cloud-level \citep{Hubbard1999,Kaspi2010a,Kaspi2013a},
with the underlying assumption of geostrophic balance for low Rossby
number flows \citep{Pedlosky1987}. This leads to thermal wind balance
between the flow and the density fields. These density perturbations
will manifest in the latitudinal variations of the gravity field.
Subsequent studies examined in more detail the flow-density relation
\citep{Zhang2015,Kaspi2016,Cao-Stevenson-2017b,Galanti2017a}.

The flow field within Jupiter has the potential of being further constrained
by the planetary magnetic field. The strong magnetic field of Jupiter
(around $6$~Gauss at the surface) could affect the flow field at
depth with modestly high electrical conductivity \citep{Liu2008,heimpel2011,Gastine2014b,Jones2014,Connerney2017}.
Using models based on electrical conductivity estimates and comparing
the wind-induced Ohmic dissipation to the observed planetary luminosity,
\citet{Liu2008} estimated that the measured magnetic field strength
limits the maximum depth to which fast zonal flows can penetrate to
$0.96$ of the radius for Jupiter. \citet{Glatzmaier2008} argued
that this depth could be an overestimate due to the possible geometrical
alignment between the deep zonal flow and magnetic fields. Other studies
found that the flow itself could alter the magnetic field in the semiconducting
region \citep{heimpel2011,Gastine2014b,Cao2017}. This modification
could be detectable by the Juno magnetic field experiments \citep{Cao2017}.

\begin{figure}
\centering{}\includegraphics[scale=0.4]{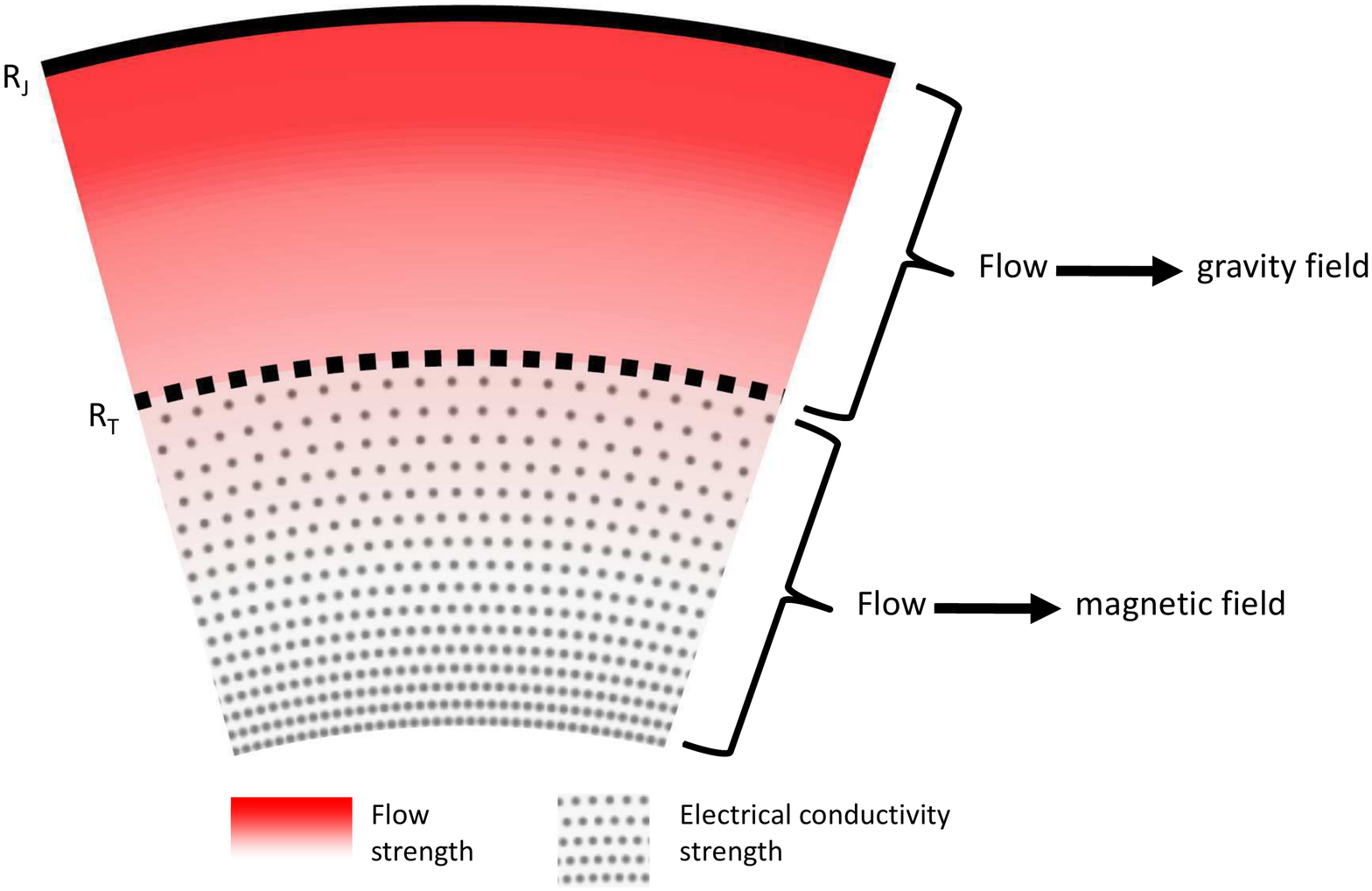}\caption{\label{fig:concept} A conceptual look at the two regions affecting
the anomalous gravity and magnetic fields. An upper region between
the planet surface $R_{J}$ and a transition level $R_{T}$, and a
lower region below that level. Shown are the strength of the flow
(from red to white), where solid red is the wind at the cloud-level
and white denotes a few order of magnitude smaller values, and the
strength of the electrical conductivity (gray dots where dense dots
represent high electrical conductivity and spare dots represents low
electrical conductivity ).}
\end{figure}

The two distinctly different dynamical regimes inside Jupiter are
conceptually presented in Fig.~\ref{fig:concept}. Starting from
the planet's upper surface, $R_{J}$, the measured cloud-level wind
is assumed to decay toward the interior (fading red color). This flow
pattern, when translated to density perturbations via the thermal
wind balance \citep[e.g.,][]{Kaspi2010a,Galanti2016}, has an integrated
signature on the gravity field. This effect has a strong depth dependency
- a density perturbation with the same magnitude will have a larger
effect on the gravity field when located closer to the planet's surface.
Alongside this dynamical effect, there exists the interaction between
the flow and the magnetic field. Below a transition depth $R_{T}$,
the electrical conductivity (grained area) has increased to a large
enough value so the flow can generate sensible magnetic perturbations
\citep{Cao2017}. This results in an anomalous latitude-dependent
magnetic field. Thus, the interior flow has two distinct regions of
influence - an upper region (between $R_{J}$ and $R_{T}$) in which
the flow affects mostly the gravity field, and a lower region (below
$R_{T}$) in which the flow affects mostly the magnetic field. To
date, there has been no study that couples the flow-magnetic balance
and the flow-gravity balance to constrain the flow field in a unified
approach. In this study we address this issue, and propose a new method
for using the expected Juno measurements of both the gravity and magnetic
fields in order to constrain the depth dependent flow field inside
Jupiter.

The manuscript is organized as follows: in section~\ref{sec:Methodology}
we describe the mean-field electrodynamics (MFED) and thermal wind
(TW) models, and present the experimental setup. In section~\ref{sec:Results}
we discuss the optimized solutions of the flow in the MFED and TW
models, and the results of the optimization when a unified decay function
is used. We conclude in section~\ref{sec:Discussion-and-conclusions}.

\section{Methodology\label{sec:Methodology}}

\subsection{The mean-field electrodynamic (MFED) balance\label{subsec:The-mean-field-electrodynamics}}

The model used here is based on the study of \citet{Cao2017}. The
steady state balance between the anomalous magnetic field and the
flow is
\begin{eqnarray}
\eta_{E}\left(\nabla^{2}-\frac{1}{s^{2}}\right)B+\frac{1}{r}\frac{d\eta_{E}}{dr}\frac{\partial(rB)}{\partial r} & = & -{\bf B_{0}}\cdot{\bf \nabla}U,\\
\eta_{E}\left(\nabla^{2}-\frac{1}{s^{2}}\right)A & = & -\alpha B,\label{eq:MHD-equations}
\end{eqnarray}
where $A(r,\theta,t)$ and $B(r,\theta,t)$ compose the anomalous
magnetic field $\mathbf{B}=\nabla\times(A\hat{e}_{\phi})+B\hat{e}_{\phi}$,
$\eta_{E}(r)$ is the effective magnetic diffusivity, $s=r\sin\theta$
is the distance from the axis of rotation, $\mathbf{B_{0}}(r,\theta)$
is the background planetary magnetic field, $\alpha(r,\theta)$ is
the dynamo $\alpha$-effect, and $U(r,\theta)$ is the zonal flow.
The magnetic diffusivity is inversely proportional to the electrical
conductivity $\sigma(r)$. The background magnetic field $\mathbf{B_{0}}=B_{0}^{r}\hat{e}_{r}+B_{0}^{\theta}\hat{e}_{\theta}$
is defined as a dipole field with $B_{0}^{r}(\theta,r)=2g_{1}^{0}r^{-3}\cos(\theta),$
and $B_{0}^{\theta}(\theta,r)=g_{1}^{0}r^{-3}\sin(\theta),$ where
$g_{1}^{0}$ is the dipole Gauss coefficient. Note that including
quadrupole and octuple moments in the background magnetic field ($\mathbf{B}_{0})$
has negligible effect on the results in this study (see Supporting
Information).

The flow $U(r,\theta)$ is defined as
\begin{eqnarray}
U(r,\theta) & = & U_{{\rm surf}}(s)Q_{M}(r),\label{eq:MHD-flow}
\end{eqnarray}
\begin{eqnarray}
Q_{M}(r) & = & \left[1+(f_{M}-1)\right]\left(\frac{r-R_{J}}{R_{T}-R_{J}}\right)^{D},\quad r>R_{T}\label{eq:MHD-decay-function-outer}
\end{eqnarray}

\begin{eqnarray}
Q_{M}(r) & = & f_{M}\exp\left(\frac{r-R_{T}}{H_{M}}\right),\quad r\leq R_{T}\label{eq:MHD-decay-function-inner}
\end{eqnarray}
where $U_{{\rm surf}}(s)$ is the measured cloud-level wind projected
towards the planet interior parallel to the spin-axis (hence its dependance
on $s$), $R_{J}$ is the planetary radius, $R_{T}$ is the transition
depth set to $0.972R_{J}$, $f_{M}$ is the ratio between the flow
strength at the transition depth and the flow at the cloud-level,
$H_{M}$ is the decay scale-height in the inner layer, and {\large{}$D=\frac{(R_{J}-R_{T})f_{M}}{(1-f_{M})H_{M}}$}
ensures the smoothness of wind across the transition depth. The measured
wind $U_{{\rm surf}}(s)$ used here is based on \citet{Porco2003}
(e.g., see \citet{Galanti2016}, Fig.~1a).

This functional form of the flow's radial decay allows two distinctly
different behaviors in the two layers. In the outer layer (Eq.~\ref{eq:MHD-decay-function-outer}),
the decay function represents a non-magnetic dynamical effect such
as the baroclinic thermal wind effect \citep{Kaspi2009}, with the
free parameter $H_{M}$ allowing a wide range of flow structures.
In the inner layer (Eq.~\ref{eq:MHD-decay-function-inner}), the
exponential decay function is assumed to be a result of the increased
electrical conductivity $\sigma$ (shown in Fig.~\ref{fig:U_decay_MHD_TW},
dashed blue lines). The strength of the electrical conductively, controls
the effect of the flow $U$ on the anomalous magnetic field ${\bf B}$.

There are two physical reasons for choosing the transition
depth to be at $0.972$ $R_{J}$. First, the depth at which the Lorentz
force could balance the observed surface Reynolds stress is around
0.972 $R_{J}$. Second, projecting the observed equatorial super-rotation
along the spin-axis towards the deep interior of Jupiter, its deepest
point is at 0.972 $R_{J}$. Furthermore, the possibility of a dynamo
layer close to 0.972 $R_{J}$ can be safely ruled out given our understanding
of the electrical conductivity of hydrogen inside Jupiter \citep{french2012}.

\subsubsection{Definition of the MFED optimization}

Juno will measure the poloidal component of the magnetic field $\mathbf{B^{\mathbf{Juno}}}=\nabla\times(A\hat{e}_{\phi})$
at the location of the spacecraft. These measurements can then be
projected to a radial location inside the planet in which the electrical
conductivity is negligible, using potential field continuation. Given
the very small electric conductivity above the transition radius $R_{T}$,
it is natural to choose $R_{T}$ as the depth of comparison. Therefore,
the 'measurements' to be used in our model are the downward continuation
of both $B_{r}^{{\rm Juno}}$ and $B_{\theta}^{{\rm Juno}}$ to the
depth $R_{T}$,
\begin{eqnarray}
B_{r}^{{\rm obs}} & = & B_{r}^{{\rm Juno}}(r=R_{T}),\nonumber \\
B_{\theta}^{{\rm obs}} & = & B_{\theta}^{{\rm Juno}}(r=R_{T}).\label{eq:obs-MHD}
\end{eqnarray}
 Our goal is to find the flow structure that will results in an anomalous
magnetic field that best matches the measured one. Therefore, a scalar
measure (a cost function, see \citealt{Galanti2016,Galanti2017b})
for the difference between the measured and the model magnetic fields
is to be defined. We set it as the difference between the measurements
and the model solution at $R_{T}$, integrated over latitude
\begin{eqnarray}
L & = & \frac{1}{\pi}\intop_{\theta=-\pi/2}^{\pi/2}\left[W_{r}^{M}(\theta)\left(B_{r}^{{\rm obs}}-B_{r}^{{\rm mod}}\right)^{2}+W_{\theta}^{M}(\theta)\left(B_{\theta}^{{\rm obs}}-B_{\theta}^{{\rm mod}}\right)^{2}\right]d\theta,\label{eq:cost-MHD}
\end{eqnarray}
where $B_{r}^{{\rm mod}}$ and $B_{\theta}^{{\rm mod}}$ are the latitudinal
dependent model solutions, and $W_{r}^{M}(\theta),W_{\theta}^{M}(\theta)$
are the weights reflecting the uncertainties in the measurements.
A gross estimate of the measurements uncertainty due to instruments
limitations is $0.01$\% of the field strength (Jack Connerney, private
communication). Given that the measured background dipole field of
Jupiter is about $4\times10^{5}$~nT, the measurements error in the
model is $B_{{\rm err}}=40$~nT. The weights in the cost function
are then
\begin{eqnarray}
W_{r}^{M},W_{\theta}^{M} & \equiv & \frac{1}{(B_{{\rm err}})^{2}}=16\cdot10^{8}.\label{eq:weights-MHD}
\end{eqnarray}

Two parameters control the flow structure in the MFED model, the scale
depth of the flow in the inner layer $H_{M}$ and the relative strength
of the wind at the transition depth $f_{M}$. Searching for the values
of these two parameters (that will result in a magnetic field that
best matches the measurements) is an optimization problem that can
be solved with various techniques. Here we follow the methodology
used recently for optimization of the TW model \citep{Galanti2016,Galanti2017b,Galanti2017c},
in which a solution is searched for iteratively, within the range
of $10<H_{M}<5000$~km and $0.025<f_{M}<1$.

\subsection{The thermal wind (TW) balance\label{subsec:The-thermal-wind}}

Similar to the MFED model, the flow field used in the thermal wind
model is based on cloud-level wind, projected into the interior on
cylinders parallel to the axis of rotation and is assumed to decay
in the radial direction. In the TW model, the decay profile used in
earlier studies \citep[e.g.,][]{Galanti2016,Galanti2017b} was based
on an exponential function. This choice might be replaced with any
other function that assures a decay of the flow in the radial direction.
Here we use a hyperbolic tangent function
\begin{eqnarray}
U_{T}(r,\theta) & = & U_{{\rm surf}}(s)Q_{T}(r),\label{eq:TW-flow}
\end{eqnarray}
\begin{eqnarray}
Q_{T}(r) & = & \frac{\tanh\left(-\frac{R_{J}-H_{T}-r}{\Delta H_{T}}\right)+1}{\tanh\left(\frac{H_{T}}{\Delta H_{T}}\right)+1},\label{eq:TW-decay-function-1}
\end{eqnarray}
where $H_{T}$ is the decay depth and $\Delta H_{T}$ is the width
of the decay. Note that the hyperbolic function is normalized by its
value at the surface of the planet to assure that the surface flow
has the value of the measured cloud-level wind. We choose this function
in order to allow a better compatibility with the function used in
the MFED model (see more details in section~\ref{subsec:The-radial-decay}).

Since the planet is rapidly rotating, and we are considering large
scale flows with small Rossby number, the flow is assumed to be geostrophic
and in thermal wind balance \citep[e.g.,][]{Kaspi2009,Kaspi2010a}.
Given that only the zonal mean flow in the azimuthal direction is
considered, the balance is
\begin{equation}
2\Omega\frac{\partial}{\partial z}\left(\widetilde{\rho}U\right)=g_{0}\frac{\partial}{\partial\theta}\rho',\label{eq: thermal wind}
\end{equation}
where $\Omega$ is the planetary rotation rate, $\widetilde{\rho}(r)$
is the background density field, $g_{0}\left(r\right)$ is the mean
gravity acceleration in the radial direction, and $\rho'\left(r,\theta\right)$
is the density anomaly associated with the flow field. The equation
is solved similarly to the method used in \citet{Galanti2017b} and
\citet{Galanti2017c}. The gravity moments (to be compared with the
measurements) are then integrated from the density field
\begin{equation}
\Delta J_{n}^{{\rm mod}}=-\frac{2\pi}{MR_{J}^{n}}\intop_{0}^{R_{J}}r{}^{n+2}dr\intop_{\mu=-1}^{1}P_{n}\left(\mu\right)\rho'\left(r,\mu\right)d\mu,\label{eq: Jn model}
\end{equation}
where $\Delta J_{n}^{{\rm mod}},\,n=2,...,N$ are the coefficients
of the gravity moments, $M$ is the planet mass, $P_{n}$ are the
Legendre polynomials, and $\mu=\cos(\theta)$.

While the measured gravity field will be in terms of the gravity moments,
here we show the results in terms of the actual latitude dependent
gravity anomalies in the radial direction 
\begin{eqnarray}
\Delta g_{r}^{{\rm mod}}(\theta) & = & -\frac{GM}{R_{J}^{2}}\sum_{n}\left(n+1\right)P_{n}\left(\cos\theta\right)\Delta J_{n}^{{\rm mod}},\label{eq:gravity-latitude}
\end{eqnarray}
where $G$ is the gravitational constant \citep{Kaspi2010a}. Note
that the gravity field in the real space is completely equivalent
to the gravity field in the moment space.

\subsubsection{Definition of the TW optimization}

The Juno measurements to be used to optimize the TW model are the
gravity moments \citep{Finocchiaro2010} from which the static body
moments are subtracted
\begin{equation}
\Delta J_{n}^{{\rm obs}}=J_{n}^{{\rm obs}}-J_{n}^{{\rm solid}},\label{eq:obs-TW}
\end{equation}
where $J_{n}^{{\rm obs}}$ are the measured gravity moments, and $J_{n}^{{\rm solid}}$
are the static body solutions to be taken from a model solution \citep[e.g., ][]{Wahl2017}.
Note that for this analysis we will examine the measured gravity as
function of latitude $\Delta g_{r}^{{\rm obs}}(\theta)$ calculated
similarly to the model solution (Eq.~\ref{eq:gravity-latitude}). 

The cost function, measuring the fit of the model solution to the
measurements, can be written in matrix notation as
\begin{eqnarray}
L & = & \left(\Delta J^{{\rm obs}}-\Delta J^{{\rm mod}}\right)^{T}W_{T}\left(\Delta J^{{\rm obs}}-\Delta J^{{\rm mod}}\right),\label{eq:cost-TW}
\end{eqnarray}
where $\left(\Delta J^{{\rm obs}}-\Delta J^{{\rm mod}}\right)$ is
a vector of the differences between the measurements and the model
solution, and $W_{T}$ is a weight matrix to be derived from the covariance
matrix of the measurements \citep{Finocchiaro2010,Galanti2017c}.
The parameters to be optimized are the decay depth $H_{T}$ and the
width of the decay $\Delta H_{T}$, both of which determine the shape
of the decay function (Eq.~\ref{eq:TW-flow}). The range of the parameters
allowed during the optimization is between $100$ and $5000$~km
for both parameters.

\subsection{Setup of the simulated flow field\label{subsec:The-radial-decay}\label{subsec:Setup-of-the-decay-profile}}

Until the Juno measurements become available, we can simulate a flow
structure that mimics a possible realistic scenario, and use the MFED
and TW models to search for the best fit solutions (see details on
methodology in \citealt{Galanti2016,Galanti2017b,Galanti2017c}).
In both the MFED and TW models, the cloud-level wind is extended inward
as a function of the distance from the axis of rotation based on angular
momentum considerations, and the wind velocity decays in the radial
direction toward the center of the planet (Eqs.~\ref{eq:MHD-decay-function-outer},\ref{eq:MHD-decay-function-inner}
and Eq.~\ref{eq:TW-flow}, respectively). The difference between
the two models is in the specifics of the radial decay profile. Since
the functions are very different from each other, the decay profile
to be used for the simulation of the measurements should be such that
neither of the two models will be able to fit it perfectly.

\begin{figure}[h]
\centering{}\includegraphics[scale=0.3]{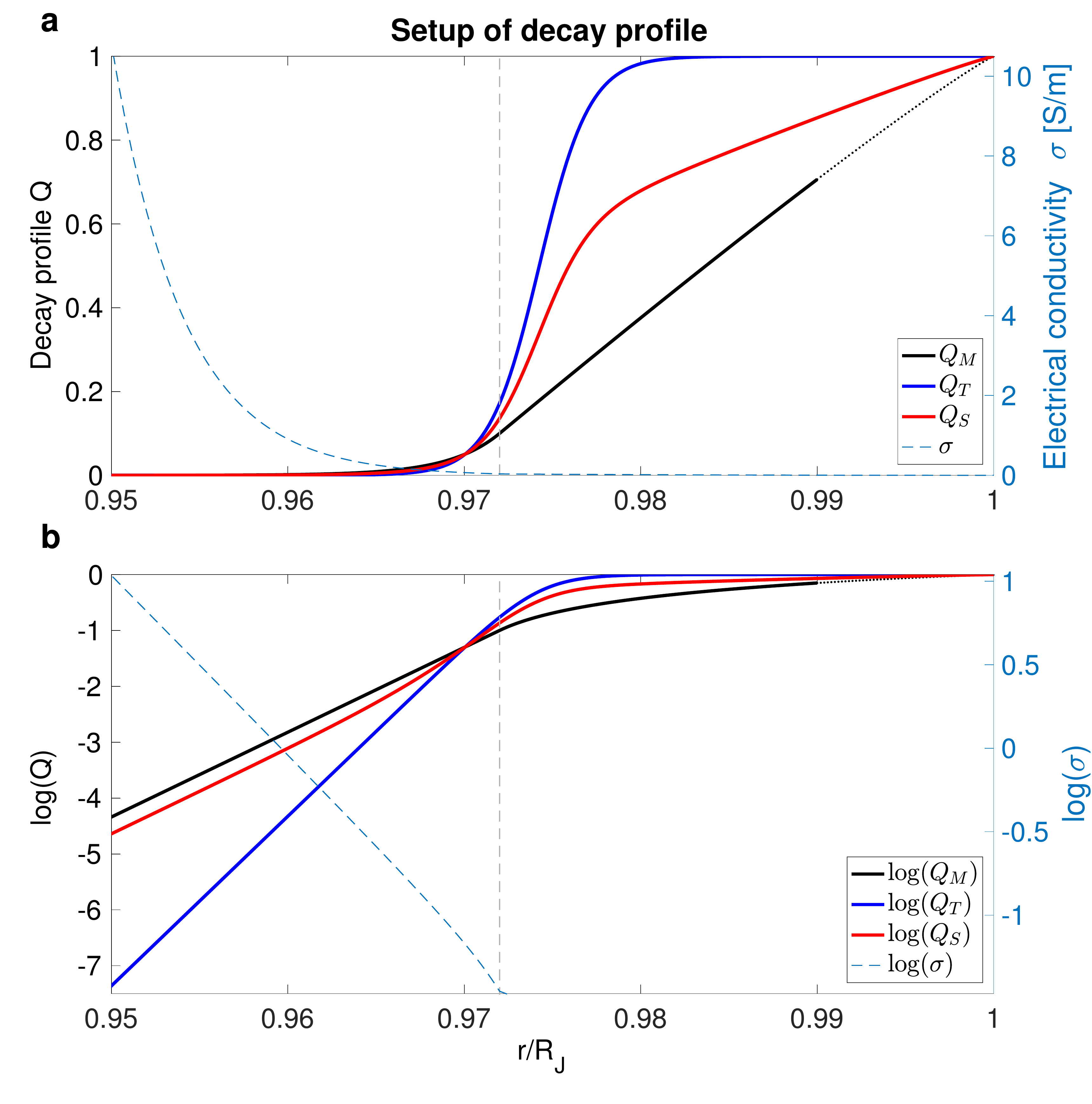}\caption{\label{fig:U_decay_MHD_TW} (a) A MFED decay factor $Q_{M}$ (black),
a TW decay factor $Q_{T}$ (blue), and their average $Q_{S}$ (red).
The solid black line is the radial range in which the MFED model is
solved. Also shown is the electrical conductivity $\sigma$ in the
MFED model, which controls the strength of the interaction between
the flow field and the magnetic field. (b) The log of the fields shown
in (a), illustrating the exponential nature (straight lines) of the
decay in the inner layer in both functions.}
\end{figure}

In Fig.~\ref{fig:U_decay_MHD_TW} we show examples of decay functions
for the MFED model (black line), the TW model (blue line), and the
average between the two models (red line). Fig.~\ref{fig:U_decay_MHD_TW}b
illustrates the exponential nature of the decay in the inner region
in both functions. The decay functions were chosen such that at the
depth of the transition to the semi conducting region $R_{T}$ the
MFED function has a value of $0.1$ and the TW a value of $0.2$.
The parameters for the MFED functions are $H_{M}=200$~km and $f_{M}=0.1$,
and for the TW function $H_{T}=1800$~km and $\Delta H_{T}=200$~km.
These specific profiles are chosen so that their average is different
from the separate profile. We will therefore use the averaged profile
($Q_{S}$, red line) to set the simulated flow, thus neither model
alone can reconstruct the flow field well. Based on this flow structure,
the 'measured' anomalous magnetic field $B_{r}^{{\rm obs}},B_{\theta}^{{\rm obs}}$
and the anomalous gravity field $\delta g_{r}^{{\rm obs}}$ are calculated
and shown in Fig.~\ref{fig:MHD_TW_obs_sol}a and Fig.~\ref{fig:MHD_TW_obs_sol}b,
respectively (red lines). 

\section{Results\label{sec:Results}}

Here we present the results from the optimization of the two models
seperately, as well as an experiment in which the results from the
MFED optimization are used to better constrain the TW model. Note
that in all cases, due to the small number of parameters to be optimized,
the optimization procedure does not depend on the initial guess of
the parameters and that a global minimum is always reached.

\subsection{Optimizing the MFED model using magnetic field measurements only\label{subsec:Optimizing-the-MHD}}

\begin{figure}[h]
\centering{}\includegraphics[scale=0.3]{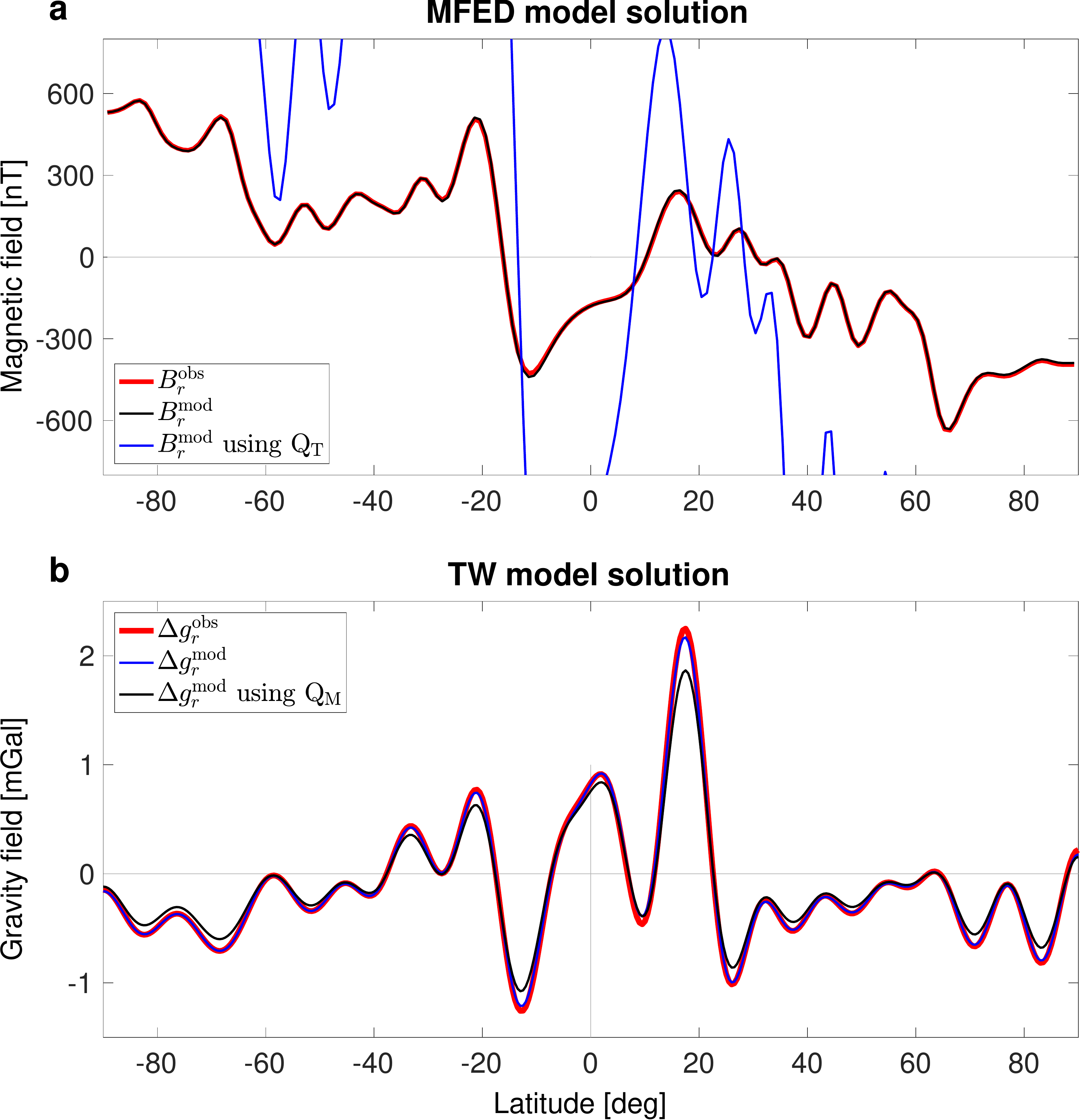}\caption{\label{fig:MHD_TW_obs_sol} (a) The simulated magnetic field (red),
and the model optimized solutions (black). (b) The simulated gravity
field (red), and the TW model optimized solution (red). Also shown
are solutions when the decay profile is taken from the other model
solution (blue lines).}
\end{figure}

Modifying the values of the parameters $H_{M}$ and $f_{M}$, an optimal
solution is reached such that the model calculated magnetic field
fits best the simulated one. A solution is reached with $H_{M}=168$~km
and $f_{M}=0.129$. Fig.~\ref{fig:MHD_TW_obs_sol}a shows the simulated
radial component of the magnetic field (red line) together with the
optimized model calculated field (black line). The match between the
solutions is excellent, with a difference that is barely distinguishable
on a linear scale. This behavior is underlined in the profile of the
optimized decay function (Fig.~\ref{fig:U_decay_solutions}a, black
line), which is in a very good agreement with the simulated one (Fig.~\ref{fig:U_decay_solutions}a,
red line) in the inner region, below $R_{T}$. On the other hand,
there is less agreement in the region between $R_{J}$ and $R_{T}$.
This implies that the MFED model is very sensitive below $R_{T}$
but less so above this depth. This will become more evident when we
discuss the case in which the two model solutions are cross-evaluated
(section~\ref{subsec:Crossing-solutions}).

\subsection{Optimizing the TW model using gravity field measurements only\label{subsec:Optimizing-the-TW}}

\begin{figure}[h]
\centering{}\includegraphics[scale=0.3]{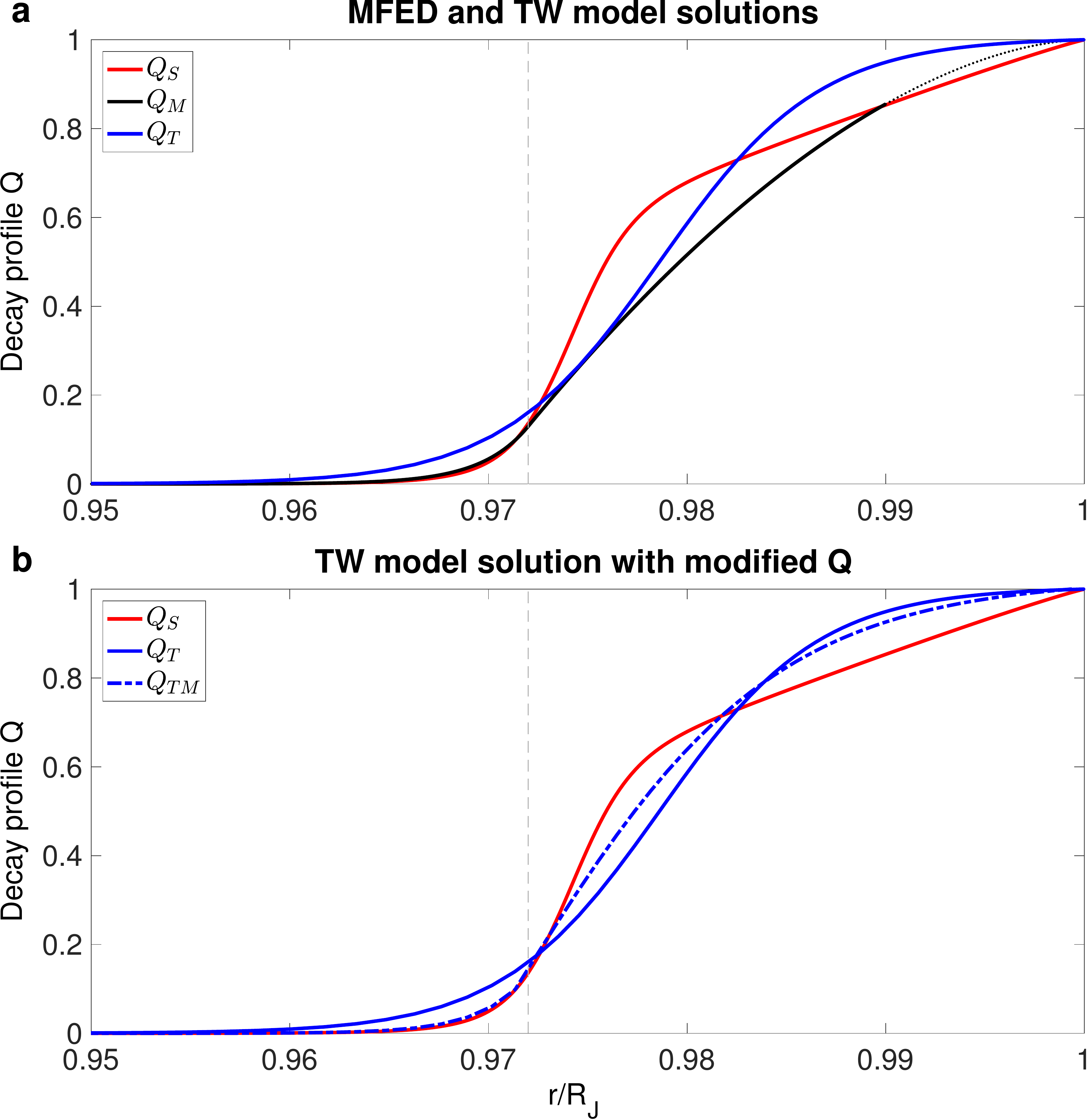}\caption{\label{fig:U_decay_solutions} (a) The radial decay factor used to
simulate the flow (red), the MFED solution (black), and the TW solution
(blow). For the MFED solution, the solid black line denote the radial
range in which the magnetic field is solved for. (b) The improved
TW solution (dash-dotted blue line) shown together with the simulation
(red) and the standard TW solution (solid blue).}
\end{figure}

Next, we optimize the values of the parameters $H_{T}$ and $\delta H_{T}$
that gives a TW model solution of the gravity field that best fit
the simulated one. A solution is reached with $H_{T}=1504$~km $\delta H_{T}=543$~km.
Fig.~\ref{fig:MHD_TW_obs_sol}b shows the simulated anomalous gravity
field (red line) together with the optimized model calculated field
(blue line). Similar to the case with the MFED model alone, the match
between the simulation and the model solution is remarkably good,
with only minor differences, for example around latitudes $17^{\circ}$S
and $17^{\circ}$N. This fit is not trivial given that the profile
of the optimized decay function (Fig.~\ref{fig:U_decay_solutions}a,
blue line) is overestimating in the region below $R_{T}$, underestimating
in the region between $R_{T}$ and around $r=0.98R_{J}$, and again
overestimating above that depth. However, the good agreement results
from the fact that gravity is an integration of the density over depth
(Eq.~\ref{eq: Jn model}) therefore the integrated contributions
of the density perturbations are such that their depth dependent deviations
from the simulation are somewhat cancelled. This will be better illustrated
in section~\ref{subsec:Crossing-solutions} where we examine the
case in which the two model solutions are cross-evaluated.

\subsection{Crossing the decay function solutions\label{subsec:Crossing-solutions}}

It is clear from the previous two experiments that both models can
reach a very good fit with the measurements, even though the flow
structure they find is not in full agreement with the simulated one.
Given that our main goal is to reconstruct the flow structure, it
is important to better understand this discrepancy. A simple test
for the validity of the decay profiles found by the two models, is
to switch between the optimized profiles and examine the effect on
the calculated magnetic and gravity fields. 

First, we use the TW model solution for the decay function $Q_{T}$
(Fig.~\ref{fig:U_decay_solutions}a, blue line) to calculate the
anomalous magnetic field in the MFED model. The resulting field (Fig.~\ref{fig:MHD_TW_obs_sol}a,
blue line) is about an order of magnitude stronger than the simulated
one, emphasizing the sensitivity of the MFED model to the strength
of the flow below $R_{T}.$ The overestimation of the TW solution
at that depth, while having very little influence on the gravity field,
has a much larger impact on the magnetic field. Next, we use the MFED
solution for the decay function $Q_{M}$ (Fig.~\ref{fig:U_decay_solutions}a,
black line) to calculate the gravity field in the TW model. The result
(Fig.~\ref{fig:MHD_TW_obs_sol}b, black line) is not as a good fit
to the simulation as was the TW model optimized solution. Overall,
the resulting gravity field is weaker than the simulated one, since
$Q_{M}$ is significantly underestimating the strength of the flow
almost everywhere between $R_{T}$ and $R_{J}$. The better estimation
below $R_{T}$ has little effect on the calculated gravity field.

\subsection{Optimizing the TW model using a unified decay function\label{subsec:Optimizing-improved-TW}}

Given the results of the above experiment, we now examine an estimation
of the flow structure combining both models. The MFED model captures
very well the decay of the winds below the transition depth $R_{T}$.
This solution can be used to constrain further the TW model solution
by using a modified decay function that is allowed to vary only above
$R_{T}$, and is set to have the values of the MFED solution from
that depth and inward. For simplicity, we also assume that the depth
$R_{T}$ is the inflection point of the hyperbolic tangent. The new
function to be optimized with the TW model can be approximated as
\begin{eqnarray}
Q_{TM}(r) & = & \tanh\left(-\frac{R_{T}-r}{\delta H_{T}}\right)\left[\tanh^{-1}\left(-\frac{R_{T}-R_{J}}{\delta H_{T}}\right)-f_{M}\right]+f_{M},\quad R_{T}<r<R_{J},\label{eq:TW-decay-modified}
\end{eqnarray}

\begin{eqnarray}
Q_{TM}(r) & = & f_{M}\exp\left(\frac{r-R_{T}}{H_{M}}\right),\quad r\leq R_{T}.\label{eq:TW-decay-modified-inner}
\end{eqnarray}

We can now optimize again the TW model, with the width of the hyperbolic
tangent $\delta H_{T}$ being the only parameter to be optimized.
The optimized solution is reached with $\delta H_{T}=870$~km. The
optimized profile is shown in Fig.~\ref{fig:U_decay_solutions}b
(dash-dotted blue line) together with the previous less constrained
profile (solid blue line) and the decay profile used for the simulation
(red line). In addition to the values below $R_{T}$ that are now
identical to the MFED model solution, the values above $R_{T}$ are
now in a much better agreement with the decay function used for the
simulation. In the region close to the surface of the planet, the
wind decays more, and in the region closer to the transition depth
$R_{T}$ it decay less. This new profile $Q_{TM}$ can now be used
also in the MFED model with a resulting magnetic field that is in
excellent agreement with the simulation.

\section{Conclusions\label{sec:Discussion-and-conclusions}}

A new coupled method is proposed for combining the gravity and magnetic
field measurements from the Juno mission, in order to decipher the
flow structure below Jupiter's cloud-level. We do so using a mean-field
electrodynamic balance that relates the flow field to the anomalous
magnetic field, and geostrophic balance (implying also thermal wind
balance) that relates the flow field to the anomalous gravity field.
A flow structure that fits neither model perfectly is used to simulate
the measurements, therefore posing a non-trivial problem for the optimization
of the models.

We find that the decay profile of the flow can be fit in both models
to give a very good match to the simulated measurements, even though
the two decay profile solutions differ significantly from one another.
The reason for that apparent paradox is that there are two separate
regions of influence (Fig.~\ref{fig:concept}). An upper region (between
$R_{J}$ and $R_{T}$) in which the flow is strong and in which no
interaction with the magnetic field exists, and a lower region (below
$R_{T}$) in which the flow is weak but is affecting the magnetic
field strongly.

Importantly, placing an effective constraint on the strength
of deep zonal flows in the semi-conducting region of Jupiter does
not require a fit as good as reached in this study. For example, an
absence of strongly banded magnetic field in the observations would
already impose strong constraints on the amplitude and vertical scale
height of zonal flows in the semi-conducting region. Furthermore,
separating the dynamo-generated magnetic field from the wind-induced
magnetic field is not trivial. As pointed out in \citet{Cao2017},
one way to overcome this difficulty would be to perform local inversions
of the measured magnetic field in order to extract $B_{r}$ at a limited
latitudinal band. Since we do not expect the dynamo-generated magnetic
field to feature structures at length-scales that are similar to the
surface zonal wind length scale, such local inversions could help
identify small-scale features in $B_{r}$ that are correlated with
surface winds. Note also that while here we use a simplified $\alpha$-effect, a more complex
$\alpha$-effect with latitudinal dependence, as well as the $\gamma$-effect
\citep{Kapyla2006}, are certainly possible. However, while adding
complexity to the expected solutions, including these parameters would
not change the method presented in our study.

Crossing the results from the two models, i.e., using the decay profile
solution from the TW model to calculate the magnetic field, and using
the decay profile solution from the MFED model to calculate the gravity
field, shows the discrepancies between the two approaches. The TW
profile's over-estimation of the flow amplitude at the lower region
results in a magnetic field that is an order of magnitude larger than
the simulated one, and the MFED profile's under-estimation of the
flow amplitude in the upper region results in a gravity field that
is about $20\%$ weaker than the simulated measurements. One way to
overcome the discrepancy, and therefore to utilize both approaches
combined, is offered. A new profile is set for the TW model, such
that in the lower region the solution from the MFED model is used,
and in the upper region a new profile is optimized consistently with
the lower region values.

The Juno measurements to date, have already reformed our understanding
of Jupiter's interior from the gravity \citep{Folkner2017,Kaspi2017,Wahl2017},
microwave \citep{Li2017} and magnetic measurements \citep{Connerney2017,Moore2017}.
This study presents the first unified approach for combining these
measurements, taking advantage of the sensitivity of the magnetic
and gravity fields to different depth regimes within the planet. As
the measurements become more abundant this approach may be beneficial
in building an understanding of the 3D flow within the planet. 

\textit{Acknowledgments:} This research has been supported by the
Israeli Ministry of Science and the Minerva foundation with funding
from the Federal German Ministry of Education and Research. EG and
YK also acknowledge support from the Helen Kimmel Center for Planetary
Science at the Weizmann Institute of Science. Most of the numerical
information is provided in the figures produced by solving the equations 
in the paper. Any additional data may be obtained from EG (email: eli.galanti@weizmann.ac.il)).


\end{document}